%% file: main.tex
\documentclass[runningheads]{llncs}
\usepackage[inline]{enumitem}
\usepackage{fontawesome5}
\usepackage{graphicx}
\usepackage[utf8]{inputenc} 
\usepackage[numbers]{natbib}
\usepackage{siunitx}
\usepackage[hyphens]{url}
\usepackage{tabularx}
\usepackage{tikz}
\usepackage{xcolor}

\usepackage{array}
\newcolumntype{L}[1]{>{\raggedright\let\newline\\\arraybackslash\hspace{0pt}}m{#1}}
\newcolumntype{C}[1]{>{\centering\let\newline\\\arraybackslash\hspace{0pt}}m{#1}}
\newcolumntype{R}[1]{>{\raggedleft\let\newline\\\arraybackslash\hspace{0pt}}m{#1}}

\usepackage{hyperref}
\usepackage{cleveref}

\usetikzlibrary{
    arrows,
    arrows.meta,
    calc,
    decorations.text,
    positioning,
    shapes.geometric
}

\input{colors}

\begin{document}
\title{Precise Energy Consumption Measurements of Heterogeneous Artificial Intelligence Workloads}

%
\author{
    René Caspart\inst{1} \and
    Sebastian Ziegler\inst{2} \and
    Arvid Weyrauch\inst{1} \and
    Holger Obermaier\inst{1} \and
    Simon Raffeiner\inst{1}\and
    Leon Pascal Schuhmacher\inst{1}\and
    Jan Scholtyssek\inst{2} \and
    Darya Trofimova\inst{2} \and
    Marco Nolden\inst{2} \and
    Ines Reinartz\inst{1} \and
    Fabian Isensee\inst{2} \and
    Markus Götz\inst{1} \and
    Charlotte Debus\inst{1}
}
\authorrunning{Caspart et al.}
\titlerunning{Precise Energy Consumption Measurements of Heterogeneous AI Workloads}
%
\institute{
    Karlsruhe Insitute of Technology (KIT), Germany\\
    \email{\{firstname.lastname\}@kit.edu}\and
    German Cancer Research Centre (DKFZ), Germany\\
    \email{\{firstname.lastname\}@dkfz-heidelberg.de}
}
\maketitle              
\begin{abstract}
With the rise of artificial intelligence (AI) in recent years and the subsequent increase in complexity of the applied models, the growing demand in computational resources is starting to pose a significant challenge. The need for higher compute power is being met with increasingly more potent accelerator hardware as well as the use of large and powerful compute clusters.
However, the gain in prediction accuracy from large models trained on distributed and accelerated systems ultimately comes at the price of a substantial increase in energy demand, and researchers have started questioning the environmental friendliness of such AI methods at scale. Consequently, awareness of energy efficiency plays an important role for AI model developers and hardware infrastructure operators likewise. 
The energy consumption of AI workloads depends both on the model implementation and the composition of the utilized hardware. Therefore, accurate measurements of the power draw of respective AI workflows on different types of compute nodes is key to algorithmic improvements and the design of future compute clusters and hardware.
Towards this end, we present measurements of the energy consumption of two typical applications of deep learning models on different types of heterogeneous compute nodes.
Our results indicate that 
\begin{enumerate*}
    \item contrary to common approaches, deriving energy consumption directly from runtime is not accurate, but the consumption of the compute node needs to be considered regarding its composition;
    \item neglecting accelerator hardware on mixed nodes results in overproportional inefficiency regarding energy consumption;
    \item energy consumption of model training and inference should be considered separately -- while training on GPUs outperforms all other node types regarding both runtime and energy consumption, inference on CPU nodes can be comparably efficient. 
\end{enumerate*}
One advantage of our approach is the fact that the information on energy consumption is available to all users of the supercomputer and not just those with administrator rights, enabling an easy transfer to other workloads alongside a raise in user-awareness of energy consumption.

\end{abstract}
\section{Introduction}
\label{sec:introduction}
In the past decade, artificial intelligence (AI) methods have yielded great advances in many areas of science and technology. However, growing complexity in prediction tasks is followed by an equally growing size and complexity in the AI models.
Training such large models requires an enormous amount of compute resources, as demonstrated by recent publications~\cite{brown2020language,jumper2021highly}. In addition, the development process usually includes multiple test runs and hyperparameter optimization, further increasing the needed compute time.
While modern accelerator hardware and large-scale computer clusters allow AI researchers to implement such models, the extraordinary need for electricity of these IT-infrastructures poses an increasing challenge, especially with regards to climate change.
Recent studies have therefore placed a focus not only on the predictive accuracy of modern AI models, but also on their environmental friendliness in terms of energy consumption and CO$_2$ footprint~\cite{schwartz2020green}. Yet, current efforts rely mainly on estimating electricity consumption from training and prediction (inference) runtimes~\cite{strubell2019energy}. Such approaches can only give a rough approximation and do not factor in consumption differences of specific hardware components or executed tasks. To properly gauge the gain in prediction accuracy versus the additional model complexity, as well as raise user awareness on the energy consumption of their AI applications accurate measurements of AI workload energy consumption are needed.

In conventional high-performance-computing, measuring energy consumption of computer code has been investigated thoroughly. Several studies have used either external or internal power meters for assessing the power consumption of commonly used numeric algorithms~\cite{ezzatti2019power}. 
For AI models, however, there exists little work on actual measurements of electricity consumption.

Modern deep learning models are increasingly trained on large computer clusters, where measurements via external power meters are not feasible. 
An alternative is investigating electricity draw of a single device, e.g. a single GPU via NVIDIAs management library (NVML). However, AI workloads are typically run on entire compute nodes, which host nodes with more than one accelerator device or multiple thereof, connected via a fabric. Thus, the energy consumption of the entire training pipeline cannot be precisely captured by linear scaling with the number of GPUs utilized as this would neglect the consumption of the enclosing environment of the accelerator, e.g. CPU, RAM, local disks, fabric, and so forth. Furthermore, despite the tremendous success of GPUs for deep learning applications, access to accelerator hardware is still limited, and many super-computers still host mainly CPU-only nodes.

In order to assess the energy consumption of large-scale neural network training as well as raising user awareness on the carbon footprint of extensive, and potentially inefficient, AI workloads, comprehensive, easily accessible and yet precise assessment of the nodes energy consumption is needed. However, the information on hardware power draw usually requires root access to the system and is therefore not available to common users.
Towards this end, the following study presents whole node energy measurements of two use cases representing typical deep learning applications, an image classification problem and a time series forecasting problem. Energy profiles and consumption of these workloads were evaluated in a way that is available to all users of the system. 
To highlight the differences in heterogeneous hardware compositions, model training and prediction is run on different compute node types with and without GPUs.
For all of the experiments, we limit ourselves to measuring the energy consumption in an as-is state of the worker nodes of the HPC cluster. We explicitly do not optimize the node configuration, power limits and CPU frequencies to the specific use case, to imitate the usage scenario of a typical user of an HPC system.

The remainder of the paper is organized as follows: \Cref{sec:related-work} discusses prior work on the topic of measuring power consumption of compute hardware and energy-efficient AI.  \Cref{sec:experimental-evaluation} introduces the use cases, including model architectures and datasets, as well as the compute environment and energy measurement tools utilized in the study. Results of the energy measurements are presented in \Cref{sec:results}. Finally, \Cref{sec:conclusion} discusses the found results and future studies.

\section{Related Work}
\label{sec:related-work}
\subsubsection{Power Aware Computing}

Energy Efficient HPC is an important topic for the HPC community, specifically in the light of exascale clusters. Many efforts to study and improve the overall energy efficiency of HPC clusters and corresponding aspects are coordinated and conducted as part of the ``Energy Efficient High Performance Computing Working Group''\cite{eehpc}. 

Many studies are conducted on the energy consumption of HPC systems to guide the design and develop strategies to improve the energy efficiency of an HPC clusters as a whole, e.g. \cite{pakin2016energyhpc,endrei2018energyhpc,nonaka2020energyhpc}.
Additional studies consider optimizing the energy distribution in an HPC cluster \cite{huazhe2019energyhpc,neha2018energyhpc}. These focus on improving the overall performance of a cluster, while respecting an overall power limit.
\citet{patel2020energy, Patel2020JobCO} and \citet{woong2021energy} studied the power consumption and behaviour of an HPC center across many different jobs.
Our work shares commonalities with these studies, however, we focus ourselves on AI/ML workflows and aim at providing a view of the energy consumption of typical workloads in this domain, as they are performed by users of HPC clusters on a daily basis. We explore and compare different possible usage options for these jobs on the cluster, aiming to incentivise energy efficiency considerations among the users in this domain.

Several authors have published studies on energy measurements utilizing power meters, which can be categorized into two different approaches: internal or external ones.
Among others \citet{suda2009accurate} used external power meters via clamp probes with the aim of verifying a power model for workloads.
On their own, these types of measurement are not practical, since their implementation requires substantial efforts and the approach is hardly suited for larger cluster setups, such as high-performance computing clusters.
Internal power meters can be further subdivided based on which parts of a system are measured by it.
On most nowadays available NVIDIA and AMD GPUs internal power meters are available.
These can be read out using high-level libraries and tools, such as NVML~\cite{NVML} or corresponding tools for AMD.
Using NVML to provide real-time power measurement data for GPUs has been studied and compared to a proposed power model used for predicting power demands of linear algebraic kernels on GPUs~\cite{kasichayanula2012power}.
However, utilizing libraries and tools like NVML yield only power metrics for the GPUs in a system, which makes out only one part of the energy consumption of the full system. Other components such as CPU, memory or local disk, are not taken into account with this approach.
Considering the power draw of all components of a node becomes particularly important for scientists having the choice of different compute nodes to run their computation on, e.g. CPU-only nodes and nodes also equipped with GPU accelerators.

Many system vendors are integrating internal solutions for measuring the power demand of a system, which provide important information to HPC operators.
A study to make information relying on these tools also available to users of the systems is for  example the joint HDEEM project between Bull and Technical University Dresden (TUD), which aims to provide high resolution and accurate power consumption metrics~\cite{ilsche2019energy}.
The approach is also used in production at TUD enabling users to gain information on the energy consumption of their workloads. 

\subsubsection{Energy Efficiency in AI}
Recently, awareness on the energy consumption and eco-friendliness of modern AI methods has been raised~\cite{schwartz2020green}.
Yet, there are only few reports studying the actual energy consumption of modern day AI algorithms.
In general, it is assumed that a reduction in runtime, especially for training, and/or number of parameters results in more energy efficient networks. Several authors rely on estimating power consumption based on the number of used floating point operations per second (FLOPs), e.g.~\citet{brown2020language}.
To reduce training time, authors employ approaches like pre-training and few-shot learning~\cite{devlin2018bert}. To reduce the parameter count, sparsity is extensively explored in the literature. So far, these approaches are mostly limited to inference models, i.e. pruning fully trained models to smaller sizes for deployment on low-energy (embedded) hardware, e.g. FPGAs or ASICs~\cite{montgomerie2019power}.
However, direct measurements of the entire energy consumption including all hardware components is rarely performed.
\citet{strubell2019energy} estimated electricity usage and carbon dioxide footprint of training, tuning and inference of several well-known large deep learning models. Their method is based on the runtime of these models, also factoring in the effects of hyperparameter tuning.
In an attempt to further raise awareness around the carbon emissions of machine learning methods \citet{lacoste2019quantifying} presented a \textit{Machine Learning Emissions Calculator}, that estimates the CO$_2$ emission of a given model based on the geographical location of the utilized server, the type of utilized accelerator and the overall training time of the model.

\citet{li2016evaluating} evaluated the power behavior and energy efficiency of convolutional neural networks (CNN) in commonly used deep learning frameworks on both, CPUs and GPUs, namely Intel Xeon CPUs, NVIDIA K20 and Titan X GPUs. Power draw of different CNNs were assessed via Intel’s Running Average Power Limit (RAPL) interface for CPU and VRAM~\cite{david2010rapl}, and via the NVIDIA System Management Interface for GPUs.
Our work is similar to that performed by \citet{hodak2019towards}. In their study, the authors perform measurements of the total consumed energy as well as relative CPU, GPU and other hardware contributions in a typical image recognition task. They ran training of an ImageNet-based Tensorflow benchmark on multi-GPU-Servers, comprising four 32GB NVIDIA V100 GPUs and two Intel Xeon Gold 6142 CPUs, and measured both AC and DC draw over the entire AI workload through power meters embedded in the servers power supplies as well as through NVML.



\section{Experimental Evaluation}
\label{sec:experimental-evaluation}
In order to evaluate energy consumption of different AI workloads on heterogeneous hardware nodes, we performed experimental runs of two types of deep learning applications (use cases) on different types of compute nodes on a high-performance computer cluster.

\subsection{Workloads}
\label{sec:workloads}
\begin{table}[tb]
    \caption{Summary of the computational properties of the use cases \textit{Health} and \textit{Energy}.}
    \label{tab:parameters}
    \centering
  \renewcommand{\arraystretch}{1.1}
    \begin{tabularx}{0.7\linewidth}{XlR{2cm}R{2.6cm}}
    \hline
    Use Case & Model  & Parameters & FLOPs/sample \\
    \hline
    \textit{Health} & CNN  & 20.4M & 10.14G\\
    \textit{Energy} & LSTM & 9.79K & 16.13K\\
    \hline
    \end{tabularx}
\end{table}
For the use cases, two common types of AI tasks were chosen: a computer vision classification task and a time series regression task. With the aim of measuring energy consumption of AI workloads representing typical scientific applications of deep neural networks, real-world datasets for these two tasks were selected from the research fields \textit{Health} and \textit{Energy}.
For both use cases, training and prediction runs with realistic model configurations were conducted on different types of large scale compute nodes, and the overall energy consumption was measured. Table~\ref{tab:parameters} shows a high-level summary of the computational characteristics of the used deep learning models. 

\subsubsection{Use Case \textit{Energy}}
\label{subsubsec:use-case-energy}
For the use case \textit{Energy}, we chose the task of predicting future electricity consumption (load) over a 7-day period based on historic data. In terms of AI workloads, this corresponds to a classical time-series forecasting, i.e. regression problem. 
The dataset was derived from the Western Europe Power Consumption Dataset~\cite{usecase1}, which consists of five years of load data of 15 European countries. The datasets was prepared to be continuous and complete, i.e. \texttt{NaNs} were removed and all load curves were brought to a temporal resolution of 1 hour through averaging. Samples were normalized separately for each country to the interval $[0, 1]$. Training data covers the years 2014-2017, validation and test data was taken from the years 2018 and 2019, respectively.

\begin{figure}[tb]
    \centering
    \sffamily
    \resizebox{\linewidth}{!}{%
        \input{figures/energy-lstm.tikz}
    }
    \rmfamily
    \caption{Schematic LSTM architecture for the use case \textit{Energy} to forecasting electrical load for a 7-day period.}
    \label{fig:lstm}
\end{figure}
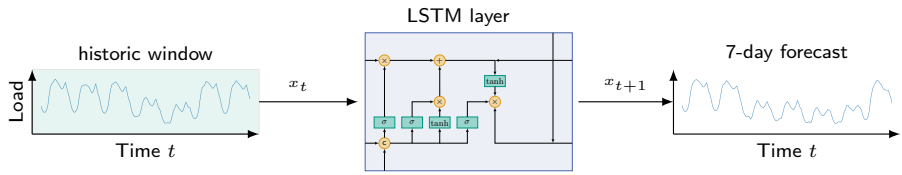

A single layer long-short term memory (LSTM) architecture (cf. \Cref{fig:lstm}) with 48 hidden nodes  was used to forecast the hourly electric demand for the next seven days based on the prior seven days load profiles as input~\cite{kong_short-term_2019,muzaffar_short-term_2019}. 
The resulting 48 output features were mapped to the required single output feature with one fully-connected layer, i.e. each recurrent loop of the model produces a one-week ahead forecast. While the model itself is rather small in terms of trainable parameters (cf. Table~\ref{tab:parameters}) the recurrence in sequence processing results in a substantial computation workload.

The model was trained for 30 epochs with the Adam optimizer at a learning rate of $10^{-3}$ and a batch size of 64. Loss was calculated as the mean squared error (MSE).
All related scripts can be found on GitHub\footnote{\url{https://github.com/Helmholtz-AI-Energy/AI-HERO-Energy}}.

\subsubsection{Use Case \textit{Health}}
\label{subsubsec:use-case-health}

The second use case \textit{Health} covered the task of predicting a COVID-19 infection based on an lung x-ray images, i.e. an image classification problem. The dataset was taken from the COVID-Net Open Initiative~\cite{covid-data} on Kaggle~\cite{usecase2}. It comprises 2$,$358 images of COVID-19-positive patients and 13,993 images of COVID-19-negative patients, collected from various sources. We employed a different data split than the one provided by Kaggle, to prevent data sources in training and test data from overlapping. The training set contains 2$,$088 positive and 13$,$696 negative samples, the validation set contains 74 positive and 76 and negative samples and the test set contains 196 positive and 221 negative samples. Images were transformed by applying a logarithmic transform and random blurring.
For the prediction model we followed the VGG-19 architecture~\cite{vgg}, adding batch normalization and replacing the three fully connected layers in the end by an average pooling and one fully connected layer.
The model was trained for 250 epochs using the SGD optimizer with a Cosine Annealing learning rate scheduler at an initial learning rate of 0.1 and a batch size of 64. Data was augmented during training by resizing, applying random horizontal flips and random rotations, taking a random crop of 224$\times$224 pixels and finally normalizing the image. For validation and testing the images only got resized to the respective size and normalized.
The entire code used to run the model can be found on GitHub\footnote{\url{https://github.com/Helmholtz-AI-Energy/AI-HERO-Health}}.

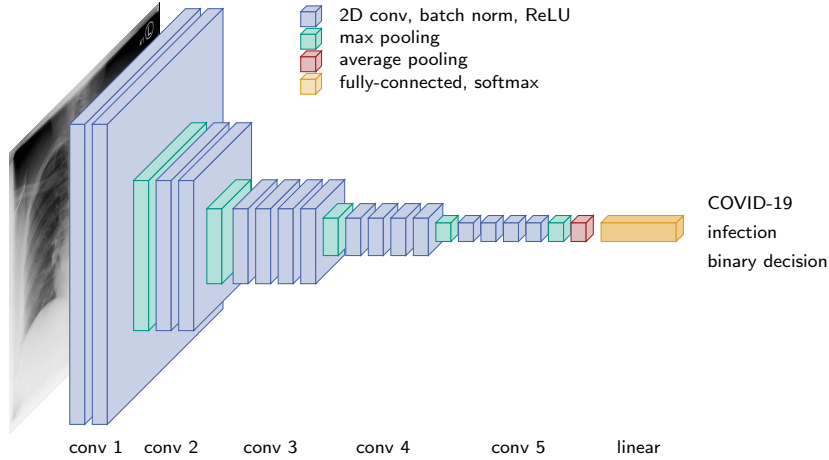
\begin{figure}[tb]
    \centering
    \resizebox{\linewidth}{!}{%
        \sffamily
        \input{figures/covid-vgg.tikz}
    }
    \caption{VGG model architecture for the use case \textit{Health} to predict a COVID-19 infection based on the x-ray input images.}
    \label{fig:vgg}
\end{figure}

\subsection{Computation Environment}
\label{sec:experimental-environment}
All experiments are conducted on the Tier-2 \textit{HoreKa} supercomputing system, an innovative hybrid cluster with nearly 60\,000 Intel Xeon ``Ice Lake'' processor cores, more than 220 terabytes of main memory, and nearly 700 NVIDIA A100 Tensor Core GPUs. The system is designed as an energy efficient system, peaking at rank 25 in the Green500~\cite{green500_0621}.
HoreKa consists of two partitions, a CPU-only partition (HoreKa-Blue) designed for highly parallel MPI applications with large memory bandwidth and an accelerated partition (HoreKa-Green) equipped with state-of-the-art accelerators for extremely data- and compute-intensive applications in machine learning.
Each of the nodes is a two socket system with Intel Xeon Platinum 8368 CPUs, 38 cores per socket, and two threads per core. It has \SI{64}{KB} L1 and \SI{1}{MB} L2 cache per core and \SI{57}{MB} shared L3 cache per CPU. Horeka-Blue nodes feature \SI{256}{GB} of main memory and one \SI{960}{GB} NVMe SSD each.
HoreKa-Green nodes are equipped with \SI{512}{GB} of main memory and four NVIDIA A100-40 GPUs.
The operating system of the nodes is Red Hat Enterprise Linux 8.2 with kernel version 4.18.0-193.60.2.el8\_2.x86\_64, with NVIDIA driver version 470.57.02, and CUDA version 11.4 for the nodes equipped with A100 accelerators.
Our use cases are implemented in Python 3.8.0 compiled with GCC 8.3.1 20191121 (Red Hat 8.3.1-5) using the PyTorch framework~\cite{paszke2019pytorch} versioned 1.11.0.dev20210929+cu111.
For the interactive access to the compute resources, we utilize the available Jupyterhub service, which uses jupyterlab 3.3.2 and jupyter\_server 1.16.0.

\subsection{Measurement Setup}

AI workloads, containing the full pipeline of either model training or inference for the two different use cases, were run as batch jobs on the HoreKa system. For measuring energy consumption, we consider four different cases of run setups, depending on the utilized hardware:

\begin{itemize}
    \itemsep 0.7em
    \item \textit{GPU}: The workload was run as exclusive on one A100 GPU of a HoreKa-Green node, while the other three GPUs were kept idle
    \item \textit{CPU-mix}: The workload was run on all 76 CPU cores of a HoreKa-Green node, while all of the four GPUs were kept idle
    \item \textit{CPU-only}: The workload was run on all 76 CPU cores of a HoreKa-Blue node, which do not contain any GPUs
    \item \textit{Jupyter}: Additionally, an entire analysis pipeline including data exploration and plotting was created in a Jupyter notebook and run on one GPU of a HoreKa-Green node. 
\end{itemize}

Energy consumption of the workloads was assessed via two different sources.
For one, internal power sensors of the HoreKa nodes were used to measure whole node energy consumption of the entire workflow. These sensors are part of Lenovo's XClarity Controller (XCC), which can be read via IPMI.
To enable access to the energy consumption information without requiring root access on the nodes or sharing of access credentials to the management interfaces of the nodes, a slurm plugin is used.
This plugin queries the information from XCC and stores it in slurm's accounting database as accumulated energy consumption.
To facilitate a reproducibility of our results and applicability of the method also to other workloads, we rely solely on information which can easily be accessed by any user of the HoreKa system.
For the evaluation, we query the average and total energy consumption for the jobs from slurm. 
As a second source of information, we utilize NVML to assess the individual energy consumption of the GPUs for the workloads \textit{GPU} and \textit{Jupyter} running on accelerator hardware. In order to profile the power draw on the GPUs, NVML was queried every \SI{500}{\milli\second}.
For statistical assessment, runs were repeated five times. We report average measurement parameters for job wall-clock time, average node power draw and overall workload energy consumption. 

\section{Results}
\label{sec:results}
\subsubsection{Use Case Energy}
The LSTM model achieved a mean absolute percentage error (MAPE) of 5.65\% on the unnormalized test dataset within the 30 epochs.
Since the test dataset is comparably small and would result in very short inference runtimes with consequently little to no noticeable energy consumption above baseline, measurements of prediction energy consumption were conducted on a separate dataset containing five copies of the training dataset. 

\begin{figure}[tb]
    \hspace{-1cm}\includegraphics[width=1.2\textwidth]{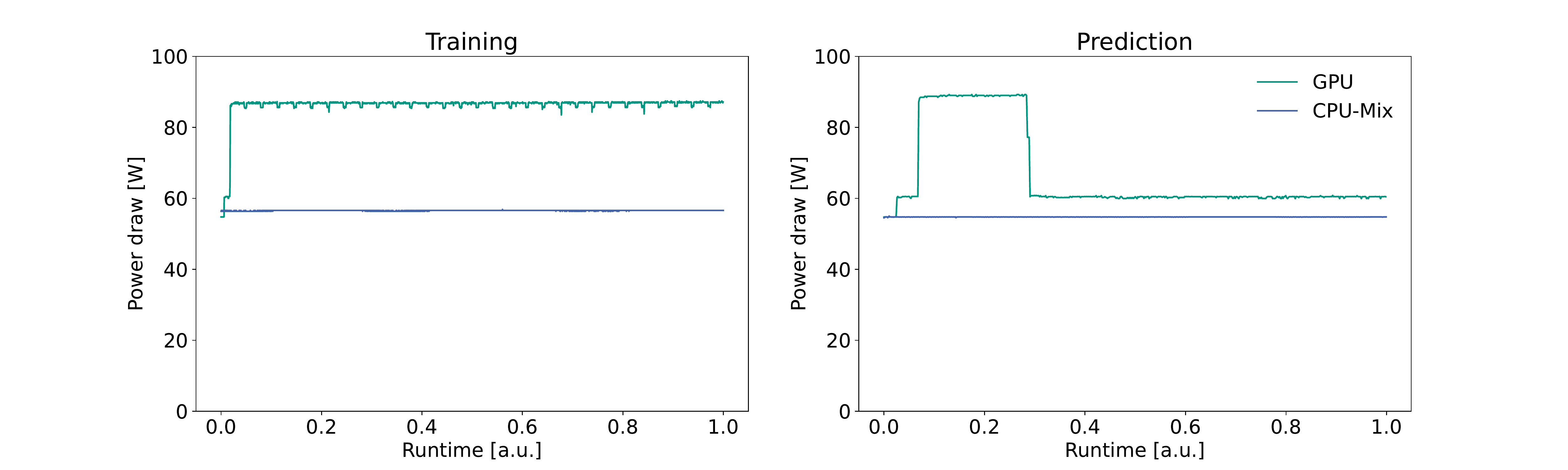}
    \caption{Jobprofile of the Energy use case, as acquired via NVML.}
    \label{fig:smi_energy}
\end{figure}

\Cref{fig:smi_energy} shows the power draw of the LSTM training and inference workload on HoreKa-Green nodes both with (green, \textit{GPU}) and without (blue, \textit{CPU-Mix}) usage of the GPU, as measured by NVML. As expected, when running the model on the nodes CPU-partition, the GPU stagnates at an idle consumption of roughly \SI{55}{\watt}. For running the model, the GPU consumes an additional energy of $\approx$ \SI{30}{\watt}, with small drops between epochs being visible. For prediction, a similar increment in energy consumption can be observed (between 0.05 and 0.3 of the fractional runtime), with the much longer low-energy idle time towards the end of the inference run attributed to result saving. 

\begin{table}[tb]
    \caption{Results of the Energy use case.}
    \label{tab:result_energy}
  \renewcommand{\arraystretch}{1.5}
    \begin{tabular}{L{2cm}L{1.6cm}R{2.8cm}R{3.8cm}R{1.8cm}}
    \hline
    & Node & Consumption [kJ]  & Average power draw [W] & Runtime\\ 
    \hline
    \textbf{Training} & GPU      & 680.7$\pm$6.7     & 665.8$\pm$9.7    & 00:17:02 \\
    & CPU-mix  & 4$\,$856.0$\pm$43.6    & 644.4$\pm$9.9   & 02:05:37 \\
    & CPU-only & 2$\,$821.5$\pm$70.8    & 374.1$\pm$11.0    & 02:05:42 \\
    \hline
    \textbf{Prediction} & GPU      & 156.9$\pm$3.4     & 606.6$\pm$8.8    & 00:04:18 \\
    & CPU-mix  & 320.2$\pm$5.9     & 621.5$\pm$7.9    & 00:08:35 \\
    & CPU-only & 189.5$\pm$4.8     & 370.6$\pm$8.7    & 00:08:31 \\
    \hline
    \textbf{Jupyter} & GPU      & 701.9    & -    & 00:17:26 \\
    \end{tabular}
\end{table}

Results of the overall node energy consumption, average power draw and runtimes of the workload on different node types are given in \Cref{tab:result_energy}. Training of the LSTM network on one NVIDIA A100 GPU is superior to running it on 76 CPU cores with respect to both runtime and energy efficiency: While \textit{GPU} runs consumed only one quarter of the energy the \textit{CPU-only} runs required, they was faster by a factor of $\approx$7.4. Although the average power draw of the \textit{GPU} runs is almost twice as much as that of the \textit{CPU-only} runs, the immense speed-up achieved through vector processing of the GPU still results in a reduced energy consumption, even for a inherently sequential problem that is a recurrent neural network. Interestingly, while runtimes on were very similar for the \textit{CPU-only} and the \textit{CPU-Mix} runs, the additional idle consumption of the GPUs on mixed nodes led to a significant increase in energy consumption by a factor of 1.7. 
Results for the inference runs however show, that even though jobs utilizing the GPUs run faster by a factor of 2, \textit{CPU-only} provides comparable energy consumption. Again, runtimes on both \textit{CPU-only} and \textit{CPU-mix} were comparable, but the additional power draw of the idle GPUs leads to a higher energy consumption of the mixed nodes.
Furthermore, we find that running a full analysis pipeline (data exploration, training and inference) in a Jupyter notebook on an A100 of the HoreKa-Green nodes results in similar energy consumption and runtimes as bash processing. However, this is under the assumption, that all cells of the notebook are executed immediately one after another, with no idle-time in between. Since this is usually not the utilization mode of Jupyter notebooks, additional baseline consumption of $\approx$ \SI{300}{W} for notebook idle time will be added for real-world applications.

\subsubsection{Use Case Health}
The VGG model of the use case \textit{Health} achieved an accuracy of 63.79\% on the test set. Training the model to full convergence (250 epochs) took \SI{2}{\hour} and \SI{34}{\minute} on one A100 GPU, with an overall energy consumption of \SI{7723.958}{\kilo\joule} and an average node power draw of \SI{835.4}{\watt}. Since running full training on CPUs of an entire node would have taken several weeks to complete, we conducted shortened experiments of 25 epochs to train the VGG model.
Due to the small size of the test dataset and the subsequent difficulties in accurately assessing inference power draw, prediction runs were modified such that each sample in the test set was used 10 times for prediction.
\begin{figure}[tb]
    \hspace{-1cm}\includegraphics[width=1.2\textwidth]{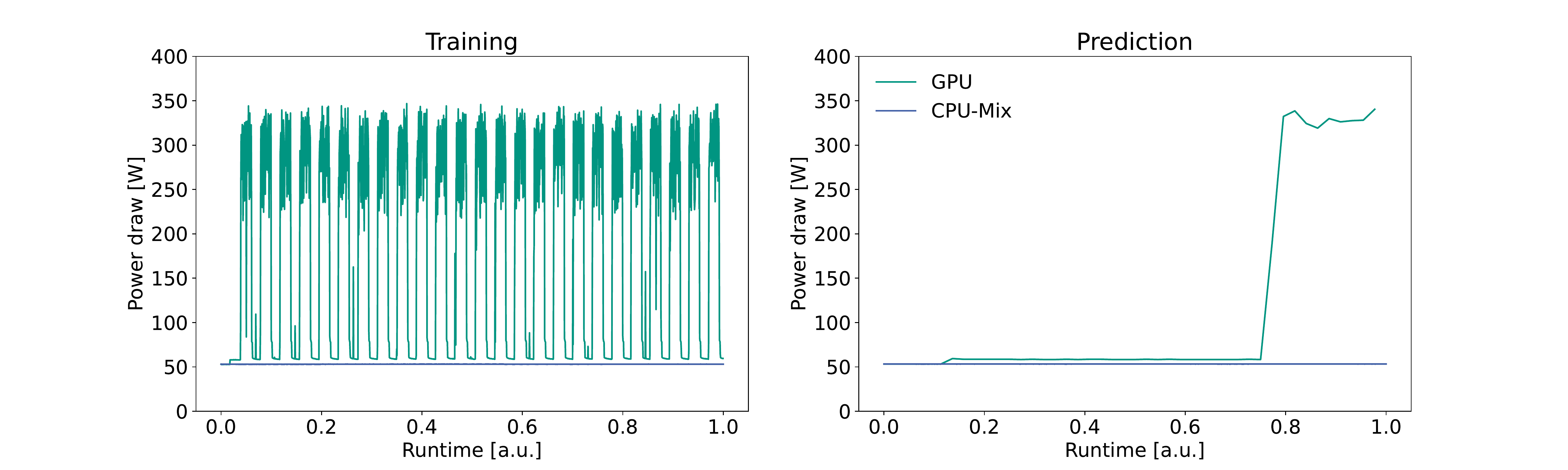}
    \caption{Jobprofile of the Health use case, as acquired via NVML.}
    \label{fig:smi_health}
\end{figure}
Results are presented in \Cref{fig:smi_health} and \Cref{tab:result_health}. The GPU power draw profile exhibits a similar behavior as previously the \textit{Energy} use case: While for the \textit{CPU-Mix} run on mixed nodes the GPU stagnates at an idle consumption around \SI{55}{\watt}, the training workload with individual epochs can be clearly seen in the \textit{GPU} run. With this use case however being much more compute intensive due to the processing of images instead of single-value time-series, the additional power draw from the workload amounts to about \SI{300}{\watt} on top of the baseline consumption. For prediction, a major fraction of the job runtime was used for data loading, which resulted only in a small increase in energy consumption. The largest contribution to the power draw budget stems from running model predictions towards the end of the workflow.

Total node energy consumption and runtime of \textit{GPU} runs is superior to runs using only CPUs in training as well as inference, even though the \textit{CPU-only} runs provide a much lower average power draw. Runs on \textit{CPU-only} require about 86 times more energy for training than those on \textit{GPU}, and 4.5 times more energy for inference.
The increase in consumed energy of runs on CPUs is not directly proportional to the increase in runtime, since prediction runs on CPUs take $\approx$ 6.3 times as long as runs on the GPU and training runs took about 194 times as long as on the GPU. Hence the electricity demand of workloads cannot safely be extrapolated from runtime alone, but there is a hardware specific component, making \textit{CPU-only} nodes still relatively efficient in terms of energy consumption. In any case, runs on \textit{CPU-mix} yielded the poorest results with respect to energy consumption as well as runtime.

Running the full training and inference pipeline in a Jupyter notebook results again in similar values for runtime and energy consumption as the batch job on a GPU. The power draw resulting from data exploration and plotting appears to be negligible in comparison to the training workload of the model. 

\begin{table}[tb]
    \caption{Results of the Health use case.}
    \label{tab:result_health}
  \renewcommand{\arraystretch}{1.3}
    \begin{tabular}{L{2cm}L{1.6cm}R{2.8cm}R{3.8cm}R{1.8cm}}
    \hline
    & Node & Consumption [kJ]  & Average power draw [W] & Runtime\\ 
    \hline
    \textbf{Training} & GPU      & 761.7$\pm$17.0     & 835.6$\pm$6.8    & 00:15:11 \\
    & CPU-mix  & 109$\,$477.1$\pm$1$\,$104.9    & 651.7$\pm$7.4   & 1-22:39:43 \\
    & CPU-only & 65$\,$796.7$\pm$4$\,$899.4    & 392.1$\pm$28.8    & 1-22:36:31 \\
    \hline
    \textbf{Prediction} & GPU      & 46.7$\pm$2.0     & 549.5$\pm$14.3    & 00:01:25 \\
    & CPU-mix  & 367.8$\pm$10.2     & 644.5$\pm$17.1   & 00:09:30 \\
    & CPU-only & 213.5$\pm$8.6     & 377.4$\pm$15.6   & 00:09:25 \\
    \hline
    \textbf{Jupyter} & GPU  & 783.6    & -    & 00:16:06 \\
    \end{tabular}
\end{table}
\section{Conclusion}
\label{sec:conclusion}

In this study, we presented high-precision measurements of whole-node energy consumption of two different AI workloads run on different heterogeneous node types of a large scale supercomputer. Our results show that for image-related deep learning models, running training and inference on a single GPU provides both shorter runtimes and lower power draw than multi-core CPU nodes. The massively parallel processing capabilities of the A100 lead to higher energy efficiency due to the significant reduction in runtime. For non-imaging workloads such as recurrent neural networks for sequential data, inference on CPU yields a comparably low energy consumption as the GPU runs, providing a valid alternative for production runs if there are no runtime constraints.

Our results further demonstrate that energy consumption of composite compute nodes cannot be estimated accurately from linear scaling in runtime of GPU consumption.  Especially for sequential data problems, a significant contribution of the energy consumption originates from the baseline of the entire node, e.g. CPU usage and memory access.

From our experiments, it is further evident that GPU idle time results in a non-negligible portion of energy consumption. Hence, GPUs should be utilized for deep learning workflows when available, even if the problem size or network architecture do not demand it straight away. This aspect also makes a strong argument for data parallel multi-GPU training, leveraging the compute power of all accelerators on a node. Finally, we showed that running AI workloads in Jupyter provides comparable energy consumption to submission via batch jobs, thereby facilitating the usage of GPUs and allowing for rapid prototyping while still maintaining energy efficiency. 

A major advantage of our approach is the fact that access to metrics of node power consumption measurements is not restricted to users with administration rights, but can be queried by every user of the system for his or her workloads. With this, AI model developers will be sensitized towards the energy footprint of their models and are able to include considerations on energy efficiency into every step of the development process.
In future studies, we aim to further map out the energy consumption of different parts of AI workflows through accurately profiling entire node power draw, as well as investigate the energy efficiency of modern AI models, namely self-attention-based architectures.
Furthermore, studies taking into account system-level optimization for the power consumption are foreseen.

\section*{Acknowledgment}
\label{sec:acknowledgment}

This work was performed on the HoreKa supercomputer funded by the Ministry of Science, Research and the Arts Baden-Württemberg and by the Federal Ministry of Education and Research.

This work is supported by the Helmholtz Association Initiative and Networking Fund under the Helmholtz AI, HIDSS4Health, Helmholtz Imaging and the Helmholtz Metadata Collaboration platform grants and the HAICORE@KIT partition.

\bibliographystyle{splncs04nat}
\bibliography{references}

\end{document}

%% file: colors.tex
\definecolor{kit-green}{RGB}{0, 150, 130}
\definecolor{kit-green100}{RGB}{0, 150, 130}
\definecolor{kit-green90}{rgb}{0.1, 0.6294, 0.5588}
\definecolor{kit-green80}{rgb}{0.2, 0.6706, 0.6078}
\definecolor{kit-green75}{rgb}{0.25, 0.6912, 0.6324}
\definecolor{kit-green70}{rgb}{0.3, 0.7118, 0.6569}
\definecolor{kit-green60}{rgb}{0.4, 0.7529, 0.7059}
\definecolor{kit-green50}{rgb}{0.5, 0.7941, 0.7549}
\definecolor{kit-green40}{rgb}{0.6, 0.8353, 0.8039}
\definecolor{kit-green30}{rgb}{0.7, 0.8765, 0.8529}
\definecolor{kit-green25}{rgb}{0.75, 0.8971, 0.8775}
\definecolor{kit-green20}{rgb}{0.8, 0.9176, 0.902}
\definecolor{kit-green15}{rgb}{0.85, 0.9382, 0.9265}
\definecolor{kit-green10}{rgb}{0.9, 0.9588, 0.951}
\definecolor{kit-green5}{rgb}{0.95, 0.9794, 0.9755}

\definecolor{kit-blue}{RGB}{70, 100, 170}
\definecolor{kit-blue100}{RGB}{70, 100, 170}
\definecolor{kit-blue90}{rgb}{0.3471, 0.4529, 0.7}
\definecolor{kit-blue80}{rgb}{0.4196, 0.5137, 0.7333}
\definecolor{kit-blue75}{rgb}{0.4559, 0.5441, 0.75}
\definecolor{kit-blue70}{rgb}{0.4922, 0.5745, 0.7667}
\definecolor{kit-blue60}{rgb}{0.5647, 0.6353, 0.8}
\definecolor{kit-blue50}{rgb}{0.6373, 0.6961, 0.8333}
\definecolor{kit-blue40}{rgb}{0.7098, 0.7569, 0.8667}
\definecolor{kit-blue30}{rgb}{0.7824, 0.8176, 0.9}
\definecolor{kit-blue25}{rgb}{0.8186, 0.848, 0.9167}
\definecolor{kit-blue20}{rgb}{0.8549, 0.8784, 0.9333}
\definecolor{kit-blue15}{rgb}{0.8912, 0.9088, 0.95}
\definecolor{kit-blue10}{rgb}{0.9275, 0.9392, 0.9667}
\definecolor{kit-blue5}{rgb}{0.9637, 0.9696, 0.9833}

\definecolor{kit-red}{RGB}{162, 34, 35}
\definecolor{kit-red100}{RGB}{162, 34, 35}
\definecolor{kit-red90}{rgb}{0.6718, 0.22, 0.2235}
\definecolor{kit-red80}{rgb}{0.7082, 0.3067, 0.3098}
\definecolor{kit-red75}{rgb}{0.7265, 0.35, 0.3529}
\definecolor{kit-red70}{rgb}{0.7447, 0.3933, 0.3961}
\definecolor{kit-red60}{rgb}{0.7812, 0.48, 0.4824}
\definecolor{kit-red50}{rgb}{0.8176, 0.5667, 0.5686}
\definecolor{kit-red40}{rgb}{0.8541, 0.6533, 0.6549}
\definecolor{kit-red30}{rgb}{0.8906, 0.74, 0.7412}
\definecolor{kit-red25}{rgb}{0.9088, 0.7833, 0.7843}
\definecolor{kit-red20}{rgb}{0.9271, 0.8267, 0.8275}
\definecolor{kit-red15}{rgb}{0.9453, 0.87, 0.8706}
\definecolor{kit-red10}{rgb}{0.9635, 0.9133, 0.9137}
\definecolor{kit-red5}{rgb}{0.9818, 0.9567, 0.9569}

\definecolor{kit-yellow}{RGB}{252, 229, 0}
\definecolor{kit-yellow100}{RGB}{252, 229, 0}
\definecolor{kit-yellow90}{rgb}{0.9894, 0.9082, 0.1}
\definecolor{kit-yellow80}{rgb}{0.9906, 0.9184, 0.2}
\definecolor{kit-yellow75}{rgb}{0.9912, 0.9235, 0.25}
\definecolor{kit-yellow70}{rgb}{0.9918, 0.9286, 0.3}
\definecolor{kit-yellow60}{rgb}{0.9929, 0.9388, 0.4}
\definecolor{kit-yellow50}{rgb}{0.9941, 0.949, 0.5}
\definecolor{kit-yellow40}{rgb}{0.9953, 0.9592, 0.6}
\definecolor{kit-yellow30}{rgb}{0.9965, 0.9694, 0.7}
\definecolor{kit-yellow25}{rgb}{0.9971, 0.9745, 0.75}
\definecolor{kit-yellow20}{rgb}{0.9976, 0.9796, 0.8}
\definecolor{kit-yellow15}{rgb}{0.9982, 0.9847, 0.85}
\definecolor{kit-yellow10}{rgb}{0.9988, 0.9898, 0.9}
\definecolor{kit-yellow5}{rgb}{0.9994, 0.9949, 0.95}

\definecolor{kit-orange}{RGB}{223, 155, 27}
\definecolor{kit-orange100}{RGB}{223, 155, 27}
\definecolor{kit-orange90}{rgb}{0.8871, 0.6471, 0.1953}
\definecolor{kit-orange80}{rgb}{0.8996, 0.6863, 0.2847}
\definecolor{kit-orange75}{rgb}{0.9059, 0.7059, 0.3294}
\definecolor{kit-orange70}{rgb}{0.9122, 0.7255, 0.3741}
\definecolor{kit-orange60}{rgb}{0.9247, 0.7647, 0.4635}
\definecolor{kit-orange50}{rgb}{0.9373, 0.8039, 0.5529}
\definecolor{kit-orange40}{rgb}{0.9498, 0.8431, 0.6424}
\definecolor{kit-orange30}{rgb}{0.9624, 0.8824, 0.7318}
\definecolor{kit-orange25}{rgb}{0.9686, 0.902, 0.7765}
\definecolor{kit-orange20}{rgb}{0.9749, 0.9216, 0.8212}
\definecolor{kit-orange15}{rgb}{0.9812, 0.9412, 0.8659}
\definecolor{kit-orange10}{rgb}{0.9875, 0.9608, 0.9106}
\definecolor{kit-orange5}{rgb}{0.9937, 0.9804, 0.9553}

\definecolor{kit-lightgreen}{RGB}{140, 182, 60}
\definecolor{kit-lightgreen100}{RGB}{140, 182, 60}
\definecolor{kit-lightgreen90}{rgb}{0.5941, 0.7424, 0.3118}
\definecolor{kit-lightgreen80}{rgb}{0.6392, 0.771, 0.3882}
\definecolor{kit-lightgreen75}{rgb}{0.6618, 0.7853, 0.4265}
\definecolor{kit-lightgreen70}{rgb}{0.6843, 0.7996, 0.4647}
\definecolor{kit-lightgreen60}{rgb}{0.7294, 0.8282, 0.5412}
\definecolor{kit-lightgreen50}{rgb}{0.7745, 0.8569, 0.6176}
\definecolor{kit-lightgreen40}{rgb}{0.8196, 0.8855, 0.6941}
\definecolor{kit-lightgreen30}{rgb}{0.8647, 0.9141, 0.7706}
\definecolor{kit-lightgreen25}{rgb}{0.8873, 0.9284, 0.8088}
\definecolor{kit-lightgreen20}{rgb}{0.9098, 0.9427, 0.8471}
\definecolor{kit-lightgreen15}{rgb}{0.9324, 0.9571, 0.8853}
\definecolor{kit-lightgreen10}{rgb}{0.9549, 0.9714, 0.9235}
\definecolor{kit-lightgreen5}{rgb}{0.9775, 0.9857, 0.9618}

\definecolor{kit-purple}{RGB}{163, 16, 124}
\definecolor{kit-purple100}{RGB}{163, 16, 124}
\definecolor{kit-purple90}{rgb}{0.6753, 0.1565, 0.5376}
\definecolor{kit-purple80}{rgb}{0.7114, 0.2502, 0.589}
\definecolor{kit-purple75}{rgb}{0.7294, 0.2971, 0.6147}
\definecolor{kit-purple70}{rgb}{0.7475, 0.3439, 0.6404}
\definecolor{kit-purple60}{rgb}{0.7835, 0.4376, 0.6918}
\definecolor{kit-purple50}{rgb}{0.8196, 0.5314, 0.7431}
\definecolor{kit-purple40}{rgb}{0.8557, 0.6251, 0.7945}
\definecolor{kit-purple30}{rgb}{0.8918, 0.7188, 0.8459}
\definecolor{kit-purple25}{rgb}{0.9098, 0.7657, 0.8716}
\definecolor{kit-purple20}{rgb}{0.9278, 0.8125, 0.8973}
\definecolor{kit-purple15}{rgb}{0.9459, 0.8594, 0.9229}
\definecolor{kit-purple10}{rgb}{0.9639, 0.9063, 0.9486}
\definecolor{kit-purple5}{rgb}{0.982, 0.9531, 0.9743}

\definecolor{kit-brown}{RGB}{167, 130, 46}
\definecolor{kit-brown100}{RGB}{167, 130, 46}
\definecolor{kit-brown90}{rgb}{0.6894, 0.5588, 0.2624}
\definecolor{kit-brown80}{rgb}{0.7239, 0.6078, 0.3443}
\definecolor{kit-brown75}{rgb}{0.7412, 0.6324, 0.3853}
\definecolor{kit-brown70}{rgb}{0.7584, 0.6569, 0.4263}
\definecolor{kit-brown60}{rgb}{0.7929, 0.7059, 0.5082}
\definecolor{kit-brown50}{rgb}{0.8275, 0.7549, 0.5902}
\definecolor{kit-brown40}{rgb}{0.862, 0.8039, 0.6722}
\definecolor{kit-brown30}{rgb}{0.8965, 0.8529, 0.7541}
\definecolor{kit-brown25}{rgb}{0.9137, 0.8775, 0.7951}
\definecolor{kit-brown20}{rgb}{0.931, 0.902, 0.8361}
\definecolor{kit-brown15}{rgb}{0.9482, 0.9265, 0.8771}
\definecolor{kit-brown10}{rgb}{0.9655, 0.951, 0.918}
\definecolor{kit-brown5}{rgb}{0.9827, 0.9755, 0.959}

\definecolor{kit-cyan}{RGB}{35, 161, 224}
\definecolor{kit-cyan100}{RGB}{35, 161, 224}
\definecolor{kit-cyan90}{rgb}{0.2235, 0.6682, 0.8906}
\definecolor{kit-cyan80}{rgb}{0.3098, 0.7051, 0.9027}
\definecolor{kit-cyan75}{rgb}{0.3529, 0.7235, 0.9088}
\definecolor{kit-cyan70}{rgb}{0.3961, 0.742, 0.9149}
\definecolor{kit-cyan60}{rgb}{0.4824, 0.7788, 0.9271}
\definecolor{kit-cyan50}{rgb}{0.5686, 0.8157, 0.9392}
\definecolor{kit-cyan40}{rgb}{0.6549, 0.8525, 0.9514}
\definecolor{kit-cyan30}{rgb}{0.7412, 0.8894, 0.9635}
\definecolor{kit-cyan25}{rgb}{0.7843, 0.9078, 0.9696}
\definecolor{kit-cyan20}{rgb}{0.8275, 0.9263, 0.9757}
\definecolor{kit-cyan15}{rgb}{0.8706, 0.9447, 0.9818}
\definecolor{kit-cyan10}{rgb}{0.9137, 0.9631, 0.9878}
\definecolor{kit-cyan5}{rgb}{0.9569, 0.9816, 0.9939}

\definecolor{kit-gray}{RGB}{0, 0, 0}
\definecolor{kit-gray100}{RGB}{0, 0, 0}
\definecolor{kit-gray90}{rgb}{0.1, 0.1, 0.1}
\definecolor{kit-gray80}{rgb}{0.2, 0.2, 0.2}
\definecolor{kit-gray75}{rgb}{0.25, 0.25, 0.25}
\definecolor{kit-gray70}{rgb}{0.3, 0.3, 0.3}
\definecolor{kit-gray60}{rgb}{0.4, 0.4, 0.4}
\definecolor{kit-gray50}{rgb}{0.5, 0.5, 0.5}
\definecolor{kit-gray40}{rgb}{0.6, 0.6, 0.6}
\definecolor{kit-gray30}{rgb}{0.7, 0.7, 0.7}
\definecolor{kit-gray25}{rgb}{0.75, 0.75, 0.75}
\definecolor{kit-gray20}{rgb}{0.8, 0.8, 0.8}
\definecolor{kit-gray15}{rgb}{0.85, 0.85, 0.85}
\definecolor{kit-gray10}{rgb}{0.9, 0.9, 0.9}
\definecolor{kit-gray5}{rgb}{0.95, 0.95, 0.95}

%% file: figures/energy-lstm.tikz
\begin{tikzpicture}
    \tikzstyle{state} = [draw=kit-orange, fill=kit-orange30, minimum width=1.2cm, minimum height=0.6cm]
    \tikzstyle{arrow} = [->, -latex]
    
    \node (data) at (-4.0, 0.0) {\includegraphics[width=2.6cm]{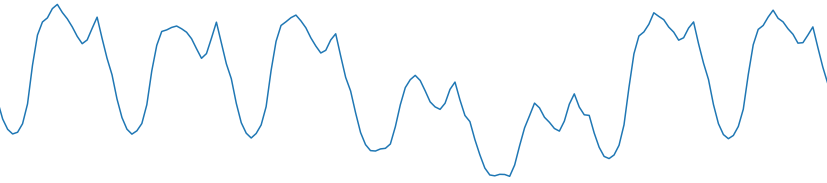}};
    \draw[fill=kit-green, opacity=0.1] (data.south west) -- (data.south east) -- (data.north east) -- (data.north west) -- cycle;
    \node[yshift=0.2cm] at (data.north) {\scriptsize historic window};
    \draw[arrow] (data.south west) -- (data.north west) node[midway, xshift=-0.2cm, rotate=90] {\scriptsize Load};
    \draw[arrow] (data.south west) -- (data.south east) node[midway, yshift=-0.2cm] {\scriptsize Time $t$};

    \node[inner sep=0, outer sep=0] (lstm1) at (-0.1, 0.0) {\resizebox{3.0cm}{!}{\input{figures/lstm_architecture.tikz}}};
    \node[yshift=0.2cm] at (lstm1.north) {\scriptsize LSTM layer};
    
    \node (label) at (4.0, 0.0) {\includegraphics[width=2.6cm]{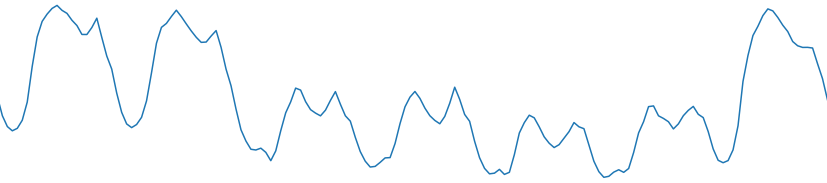}};
    \node[yshift=0.2cm] at (label.north) {\scriptsize 7-day forecast};
    \draw[arrow] (label.south west) -- (label.north west);
    \draw[arrow] (label.south west) -- (label.south east) node[midway, yshift=-0.2cm] {\scriptsize Time $t$};
    
    \draw[arrow, shorten >=-0.25cm] (data) -- (lstm1) node[midway, yshift=0.2cm] {\tiny $x_t$};
    \draw[arrow] (lstm1) -- (label) node[midway, yshift=0.2cm] {\tiny $x_{t+1}$};
\end{tikzpicture}

%% file: figures/lstm_architecture.tikz
{
    \small
    \def\edgegap{10mm}
    \def\midgap{7.5mm}
    \def\leftgap{7mm}
    
    \tikzstyle{arrow} = [->, -latex]
    \tikzstyle{rarrow} = [<-, -latex]
    \tikzstyle{cellbody} = [rectangle, minimum width = 7.5cm, minimum height = 5cm, draw=kit-blue, fill=kit-blue10]
    \tikzstyle{layer} = [rectangle, minimum width = 7.5mm, minimum height = 4mm, draw = kit-green, fill=kit-green40]
    \tikzstyle{operation} = [circle, minimum size = 4mm, inner sep = 0, draw = kit-orange, fill=kit-orange40]
    
    \begin{tikzpicture}
        \node (cell) [cellbody] {};
        \node (cat) [operation, xshift=\leftgap, yshift=\edgegap] at (cell.south west) {c};
        
        \node (times1) [operation, xshift=\leftgap, yshift=-\edgegap] at (cell.north west) {$\times$};
        \node (forgetgate) [layer, xshift=\leftgap, yshift=\edgegap+\midgap] at (cell.south west) {$\sigma$};
        
        \node (ingate1) [layer, right of=forgetgate] {$\sigma$};
        \node (ingate2) [layer, right of=ingate1] {$\tanh$};
        \node (times2) [operation, yshift=\midgap] at (ingate2) {$\times$};
        \node (add) [operation, yshift=2*\midgap] at (times2) {+};
        
        \node (outgate1) [layer, right of=ingate2] {$\sigma$};
        \node (times3) [operation, right of=outgate1, yshift=\midgap] {$\times$};
        \node (outgate2) [layer, yshift=\midgap] at (times3) {$\tanh$};
        
        \draw [arrow] (cell.north west)+(0,-\edgegap) -- (times1);
        \draw [arrow] (forgetgate) -- (times1);
        
        \draw [arrow] (cell.south west)+(0,\edgegap) -- (cat);
        \draw [arrow] (cell.south west)+(\leftgap, 0) -- (cat);
        \draw [arrow] (cat) -- (forgetgate);
        \draw [arrow] (cat) -| (ingate1);
        \draw [arrow] (ingate1)+(0,-\midgap) -| (ingate2);
        \draw [arrow] (ingate2)+(0,-\midgap) -| (outgate1);
        
        \draw [arrow] (times1) -- (add);
        \draw [arrow] (times2) -- (add);
        \draw [arrow] (ingate1) |- (times2);
        \draw [arrow] (ingate2) -- (times2);
        
        \draw [arrow] (add) -| (outgate2);
        \draw [arrow] (outgate2) -- (times3);
        \draw [arrow] (outgate1) |- (times3);
        
        \draw [rarrow] (cell.south east)+(0,\edgegap) -| (times3);
        \draw [rarrow] (cell.north east)+(0,-\edgegap) -- ([yshift=\midgap] outgate2.center);
        \draw [rarrow] (cell.north east)+(-\leftgap,0) -- ([xshift=-\leftgap, yshift=\edgegap] cell.south east);
    \end{tikzpicture}
}

%% file: figures/covid-vgg.tikz
\begin{tikzpicture}
    \node[minimum width=4.0cm, minimum height=3.75cm, yslant=1.0] at (0.0, -1.25) {\includegraphics[width=2.0cm, height=3.75cm]{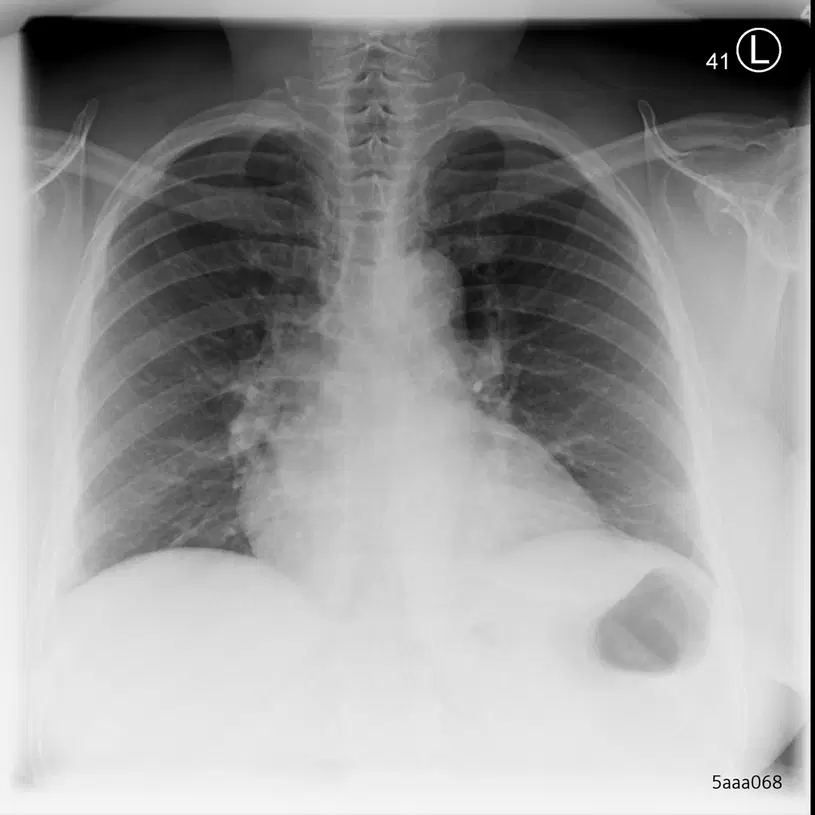}};

    \begin{scope}
        \pgfmathsetmacro{\cubex}{0.2}
        \pgfmathsetmacro{\cubey}{4}
        \pgfmathsetmacro{\cubez}{4}
        \draw[kit-blue, fill=kit-blue30] (0,0,0) -- ++(-\cubex,0,0) -- ++(0,-\cubey,0) -- ++(\cubex,0,0) -- cycle;
        \draw[kit-blue, fill=kit-blue30] (0,0,0) -- ++(0,0,-\cubez) -- ++(0,-\cubey,0) -- ++(0,0,\cubez) -- cycle;
        \draw[kit-blue, fill=kit-blue30] (0,0,0) -- ++(-\cubex,0,0) -- ++(0,0,-\cubez) -- ++(\cubex,0,0) -- cycle;
    \end{scope}
    \begin{scope}[xshift=0.3 cm]
        \pgfmathsetmacro{\cubex}{0.2}
        \pgfmathsetmacro{\cubey}{4}
        \pgfmathsetmacro{\cubez}{4}
        \draw[kit-blue, fill=kit-blue30] (0,0,0) -- ++(-\cubex,0,0) -- ++(0,-\cubey,0) -- ++(\cubex,0,0) -- cycle;
        \draw[kit-blue, fill=kit-blue30] (0,0,0) -- ++(0,0,-\cubez) -- ++(0,-\cubey,0) -- ++(0,0,\cubez) -- cycle;
        \draw[kit-blue, fill=kit-blue30] (0,0,0) -- ++(-\cubex,0,0) -- ++(0,0,-\cubez) -- ++(\cubex,0,0) -- cycle;
    \end{scope}
    \node at (0.15, -4.3) {\scriptsize conv 1};

    \begin{scope}[xshift=0.85cm, yshift=-0.75cm]
        \pgfmathsetmacro{\cubex}{0.2}
        \pgfmathsetmacro{\cubey}{2}
        \pgfmathsetmacro{\cubez}{2}
        \draw[kit-green, fill=kit-green30] (0,0,0) -- ++(-\cubex,0,0) -- ++(0,-\cubey,0) -- ++(\cubex,0,0) -- cycle;
        \draw[kit-green, fill=kit-green30] (0,0,0) -- ++(0,0,-\cubez) -- ++(0,-\cubey,0) -- ++(0,0,\cubez) -- cycle;
        \draw[kit-green, fill=kit-green30] (0,0,0) -- ++(-\cubex,0,0) -- ++(0,0,-\cubez) -- ++(\cubex,0,0) -- cycle;
    \end{scope}
    \begin{scope}[xshift=1.15cm, yshift=-0.75cm]
        \pgfmathsetmacro{\cubex}{0.2}
        \pgfmathsetmacro{\cubey}{2}
        \pgfmathsetmacro{\cubez}{2}
        \draw[kit-blue, fill=kit-blue30] (0,0,0) -- ++(-\cubex,0,0) -- ++(0,-\cubey,0) -- ++(\cubex,0,0) -- cycle;
        \draw[kit-blue, fill=kit-blue30] (0,0,0) -- ++(0,0,-\cubez) -- ++(0,-\cubey,0) -- ++(0,0,\cubez) -- cycle;
        \draw[kit-blue, fill=kit-blue30] (0,0,0) -- ++(-\cubex,0,0) -- ++(0,0,-\cubez) -- ++(\cubex,0,0) -- cycle;
    \end{scope}
    \begin{scope}[xshift=1.45cm, yshift=-0.75cm]
        \pgfmathsetmacro{\cubex}{0.2}
        \pgfmathsetmacro{\cubey}{2}
        \pgfmathsetmacro{\cubez}{2}
        \draw[kit-blue, fill=kit-blue30] (0,0,0) -- ++(-\cubex,0,0) -- ++(0,-\cubey,0) -- ++(\cubex,0,0) -- cycle;
        \draw[kit-blue, fill=kit-blue30] (0,0,0) -- ++(0,0,-\cubez) -- ++(0,-\cubey,0) -- ++(0,0,\cubez) -- cycle;
        \draw[kit-blue, fill=kit-blue30] (0,0,0) -- ++(-\cubex,0,0) -- ++(0,0,-\cubez) -- ++(\cubex,0,0) -- cycle;
    \end{scope}
    \node at (1.15, -4.3) {\scriptsize conv 2};

    \begin{scope}[xshift=1.825cm, yshift=-1.125cm]
        \pgfmathsetmacro{\cubex}{0.2}
        \pgfmathsetmacro{\cubey}{1}
        \pgfmathsetmacro{\cubez}{1}
        \draw[kit-green, fill=kit-green30] (0,0,0) -- ++(-\cubex,0,0) -- ++(0,-\cubey,0) -- ++(\cubex,0,0) -- cycle;
        \draw[kit-green, fill=kit-green30] (0,0,0) -- ++(0,0,-\cubez) -- ++(0,-\cubey,0) -- ++(0,0,\cubez) -- cycle;
        \draw[kit-green, fill=kit-green30] (0,0,0) -- ++(-\cubex,0,0) -- ++(0,0,-\cubez) -- ++(\cubex,0,0) -- cycle;
    \end{scope}
    \begin{scope}[xshift=2.175cm, yshift=-1.125cm]
        \pgfmathsetmacro{\cubex}{0.2}
        \pgfmathsetmacro{\cubey}{1}
        \pgfmathsetmacro{\cubez}{1}
        \draw[kit-blue, fill=kit-blue30] (0,0,0) -- ++(-\cubex,0,0) -- ++(0,-\cubey,0) -- ++(\cubex,0,0) -- cycle;
        \draw[kit-blue, fill=kit-blue30] (0,0,0) -- ++(0,0,-\cubez) -- ++(0,-\cubey,0) -- ++(0,0,\cubez) -- cycle;
        \draw[kit-blue, fill=kit-blue30] (0,0,0) -- ++(-\cubex,0,0) -- ++(0,0,-\cubez) -- ++(\cubex,0,0) -- cycle;
    \end{scope}
    \begin{scope}[xshift=2.475cm, yshift=-1.125cm]
        \pgfmathsetmacro{\cubex}{0.2}
        \pgfmathsetmacro{\cubey}{1}
        \pgfmathsetmacro{\cubez}{1}
        \draw[kit-blue, fill=kit-blue30] (0,0,0) -- ++(-\cubex,0,0) -- ++(0,-\cubey,0) -- ++(\cubex,0,0) -- cycle;
        \draw[kit-blue, fill=kit-blue30] (0,0,0) -- ++(0,0,-\cubez) -- ++(0,-\cubey,0) -- ++(0,0,\cubez) -- cycle;
        \draw[kit-blue, fill=kit-blue30] (0,0,0) -- ++(-\cubex,0,0) -- ++(0,0,-\cubez) -- ++(\cubex,0,0) -- cycle;
    \end{scope}
    \begin{scope}[xshift=2.775cm, yshift=-1.125cm]
        \pgfmathsetmacro{\cubex}{0.2}
        \pgfmathsetmacro{\cubey}{1}
        \pgfmathsetmacro{\cubez}{1}
        \draw[kit-blue, fill=kit-blue30] (0,0,0) -- ++(-\cubex,0,0) -- ++(0,-\cubey,0) -- ++(\cubex,0,0) -- cycle;
        \draw[kit-blue, fill=kit-blue30] (0,0,0) -- ++(0,0,-\cubez) -- ++(0,-\cubey,0) -- ++(0,0,\cubez) -- cycle;
        \draw[kit-blue, fill=kit-blue30] (0,0,0) -- ++(-\cubex,0,0) -- ++(0,0,-\cubez) -- ++(\cubex,0,0) -- cycle;
    \end{scope}
    \begin{scope}[xshift=3.075cm, yshift=-1.125cm]
        \pgfmathsetmacro{\cubex}{0.2}
        \pgfmathsetmacro{\cubey}{1}
        \pgfmathsetmacro{\cubez}{1}
        \draw[kit-blue, fill=kit-blue30] (0,0,0) -- ++(-\cubex,0,0) -- ++(0,-\cubey,0) -- ++(\cubex,0,0) -- cycle;
        \draw[kit-blue, fill=kit-blue30] (0,0,0) -- ++(0,0,-\cubez) -- ++(0,-\cubey,0) -- ++(0,0,\cubez) -- cycle;
        \draw[kit-blue, fill=kit-blue30] (0,0,0) -- ++(-\cubex,0,0) -- ++(0,0,-\cubez) -- ++(\cubex,0,0) -- cycle;
    \end{scope}
    \node at (2.475, -4.3) {\scriptsize conv 3};

    \begin{scope}[xshift=3.375cm, yshift=-1.25cm]
        \pgfmathsetmacro{\cubex}{0.2}
        \pgfmathsetmacro{\cubey}{0.5}
        \pgfmathsetmacro{\cubez}{0.5}
        \draw[kit-green, fill=kit-green30] (0,0,0) -- ++(-\cubex,0,0) -- ++(0,-\cubey,0) -- ++(\cubex,0,0) -- cycle;
        \draw[kit-green, fill=kit-green30] (0,0,0) -- ++(0,0,-\cubez) -- ++(0,-\cubey,0) -- ++(0,0,\cubez) -- cycle;
        \draw[kit-green, fill=kit-green30] (0,0,0) -- ++(-\cubex,0,0) -- ++(0,0,-\cubez) -- ++(\cubex,0,0) -- cycle;
    \end{scope}
    \begin{scope}[xshift=3.675cm, yshift=-1.25cm]
        \pgfmathsetmacro{\cubex}{0.2}
        \pgfmathsetmacro{\cubey}{0.5}
        \pgfmathsetmacro{\cubez}{0.5}
        \draw[kit-blue, fill=kit-blue30] (0,0,0) -- ++(-\cubex,0,0) -- ++(0,-\cubey,0) -- ++(\cubex,0,0) -- cycle;
        \draw[kit-blue, fill=kit-blue30] (0,0,0) -- ++(0,0,-\cubez) -- ++(0,-\cubey,0) -- ++(0,0,\cubez) -- cycle;
        \draw[kit-blue, fill=kit-blue30] (0,0,0) -- ++(-\cubex,0,0) -- ++(0,0,-\cubez) -- ++(\cubex,0,0) -- cycle;
    \end{scope}
    \begin{scope}[xshift=3.975cm, yshift=-1.25cm]
        \pgfmathsetmacro{\cubex}{0.2}
        \pgfmathsetmacro{\cubey}{0.5}
        \pgfmathsetmacro{\cubez}{0.5}
        \draw[kit-blue, fill=kit-blue30] (0,0,0) -- ++(-\cubex,0,0) -- ++(0,-\cubey,0) -- ++(\cubex,0,0) -- cycle;
        \draw[kit-blue, fill=kit-blue30] (0,0,0) -- ++(0,0,-\cubez) -- ++(0,-\cubey,0) -- ++(0,0,\cubez) -- cycle;
        \draw[kit-blue, fill=kit-blue30] (0,0,0) -- ++(-\cubex,0,0) -- ++(0,0,-\cubez) -- ++(\cubex,0,0) -- cycle;
    \end{scope}
    \begin{scope}[xshift=4.275cm, yshift=-1.25cm]
        \pgfmathsetmacro{\cubex}{0.2}
        \pgfmathsetmacro{\cubey}{0.5}
        \pgfmathsetmacro{\cubez}{0.5}
        \draw[kit-blue, fill=kit-blue30] (0,0,0) -- ++(-\cubex,0,0) -- ++(0,-\cubey,0) -- ++(\cubex,0,0) -- cycle;
        \draw[kit-blue, fill=kit-blue30] (0,0,0) -- ++(0,0,-\cubez) -- ++(0,-\cubey,0) -- ++(0,0,\cubez) -- cycle;
        \draw[kit-blue, fill=kit-blue30] (0,0,0) -- ++(-\cubex,0,0) -- ++(0,0,-\cubez) -- ++(\cubex,0,0) -- cycle;
    \end{scope}
    \begin{scope}[xshift=4.575cm, yshift=-1.25cm]
        \pgfmathsetmacro{\cubex}{0.2}
        \pgfmathsetmacro{\cubey}{0.5}
        \pgfmathsetmacro{\cubez}{0.5}
        \draw[kit-blue, fill=kit-blue30] (0,0,0) -- ++(-\cubex,0,0) -- ++(0,-\cubey,0) -- ++(\cubex,0,0) -- cycle;
        \draw[kit-blue, fill=kit-blue30] (0,0,0) -- ++(0,0,-\cubez) -- ++(0,-\cubey,0) -- ++(0,0,\cubez) -- cycle;
        \draw[kit-blue, fill=kit-blue30] (0,0,0) -- ++(-\cubex,0,0) -- ++(0,0,-\cubez) -- ++(\cubex,0,0) -- cycle;
    \end{scope}
    \node at (3.975, -4.3) {\scriptsize conv 4};

    \begin{scope}[xshift=4.875cm, yshift=-1.3125cm]
        \pgfmathsetmacro{\cubex}{0.2}
        \pgfmathsetmacro{\cubey}{0.25}
        \pgfmathsetmacro{\cubez}{0.25}
        \draw[kit-green, fill=kit-green30] (0,0,0) -- ++(-\cubex,0,0) -- ++(0,-\cubey,0) -- ++(\cubex,0,0) -- cycle;
        \draw[kit-green, fill=kit-green30] (0,0,0) -- ++(0,0,-\cubez) -- ++(0,-\cubey,0) -- ++(0,0,\cubez) -- cycle;
        \draw[kit-green, fill=kit-green30] (0,0,0) -- ++(-\cubex,0,0) -- ++(0,0,-\cubez) -- ++(\cubex,0,0) -- cycle;
    \end{scope}
    \begin{scope}[xshift=5.175cm, yshift=-1.3125cm]
        \pgfmathsetmacro{\cubex}{0.2}
        \pgfmathsetmacro{\cubey}{0.25}
        \pgfmathsetmacro{\cubez}{0.25}
        \draw[kit-blue, fill=kit-blue30] (0,0,0) -- ++(-\cubex,0,0) -- ++(0,-\cubey,0) -- ++(\cubex,0,0) -- cycle;
        \draw[kit-blue, fill=kit-blue30] (0,0,0) -- ++(0,0,-\cubez) -- ++(0,-\cubey,0) -- ++(0,0,\cubez) -- cycle;
        \draw[kit-blue, fill=kit-blue30] (0,0,0) -- ++(-\cubex,0,0) -- ++(0,0,-\cubez) -- ++(\cubex,0,0) -- cycle;
    \end{scope}
    \begin{scope}[xshift=5.475cm, yshift=-1.3125cm]
        \pgfmathsetmacro{\cubex}{0.2}
        \pgfmathsetmacro{\cubey}{0.25}
        \pgfmathsetmacro{\cubez}{0.25}
        \draw[kit-blue, fill=kit-blue30] (0,0,0) -- ++(-\cubex,0,0) -- ++(0,-\cubey,0) -- ++(\cubex,0,0) -- cycle;
        \draw[kit-blue, fill=kit-blue30] (0,0,0) -- ++(0,0,-\cubez) -- ++(0,-\cubey,0) -- ++(0,0,\cubez) -- cycle;
        \draw[kit-blue, fill=kit-blue30] (0,0,0) -- ++(-\cubex,0,0) -- ++(0,0,-\cubez) -- ++(\cubex,0,0) -- cycle;
    \end{scope}
    \begin{scope}[xshift=5.775cm, yshift=-1.3125cm]
        \pgfmathsetmacro{\cubex}{0.2}
        \pgfmathsetmacro{\cubey}{0.25}
        \pgfmathsetmacro{\cubez}{0.25}
        \draw[kit-blue, fill=kit-blue30] (0,0,0) -- ++(-\cubex,0,0) -- ++(0,-\cubey,0) -- ++(\cubex,0,0) -- cycle;
        \draw[kit-blue, fill=kit-blue30] (0,0,0) -- ++(0,0,-\cubez) -- ++(0,-\cubey,0) -- ++(0,0,\cubez) -- cycle;
        \draw[kit-blue, fill=kit-blue30] (0,0,0) -- ++(-\cubex,0,0) -- ++(0,0,-\cubez) -- ++(\cubex,0,0) -- cycle;
    \end{scope}
    \begin{scope}[xshift=6.075cm, yshift=-1.3125cm]
        \pgfmathsetmacro{\cubex}{0.2}
        \pgfmathsetmacro{\cubey}{0.25}
        \pgfmathsetmacro{\cubez}{0.25}
        \draw[kit-blue, fill=kit-blue30] (0,0,0) -- ++(-\cubex,0,0) -- ++(0,-\cubey,0) -- ++(\cubex,0,0) -- cycle;
        \draw[kit-blue, fill=kit-blue30] (0,0,0) -- ++(0,0,-\cubez) -- ++(0,-\cubey,0) -- ++(0,0,\cubez) -- cycle;
        \draw[kit-blue, fill=kit-blue30] (0,0,0) -- ++(-\cubex,0,0) -- ++(0,0,-\cubez) -- ++(\cubex,0,0) -- cycle;
    \end{scope}
    \begin{scope}[xshift=6.375cm, yshift=-1.3125cm]
        \pgfmathsetmacro{\cubex}{0.2}
        \pgfmathsetmacro{\cubey}{0.25}
        \pgfmathsetmacro{\cubez}{0.25}
        \draw[kit-green, fill=kit-green30] (0,0,0) -- ++(-\cubex,0,0) -- ++(0,-\cubey,0) -- ++(\cubex,0,0) -- cycle;
        \draw[kit-green, fill=kit-green30] (0,0,0) -- ++(0,0,-\cubez) -- ++(0,-\cubey,0) -- ++(0,0,\cubez) -- cycle;
        \draw[kit-green, fill=kit-green30] (0,0,0) -- ++(-\cubex,0,0) -- ++(0,0,-\cubez) -- ++(\cubex,0,0) -- cycle;
    \end{scope}
    \begin{scope}[xshift=6.675cm, yshift=-1.3125cm]
        \pgfmathsetmacro{\cubex}{0.2}
        \pgfmathsetmacro{\cubey}{0.25}
        \pgfmathsetmacro{\cubez}{0.25}
        \draw[kit-red, fill=kit-red30] (0,0,0) -- ++(-\cubex,0,0) -- ++(0,-\cubey,0) -- ++(\cubex,0,0) -- cycle;
        \draw[kit-red, fill=kit-red30] (0,0,0) -- ++(0,0,-\cubez) -- ++(0,-\cubey,0) -- ++(0,0,\cubez) -- cycle;
        \draw[kit-red, fill=kit-red30] (0,0,0) -- ++(-\cubex,0,0) -- ++(0,0,-\cubez) -- ++(\cubex,0,0) -- cycle;
    \end{scope}
    \node at (5.775, -4.3) {\scriptsize conv 5};

    \begin{scope}[xshift=7.875cm, yshift=-1.3125cm]
        \pgfmathsetmacro{\cubex}{1.0}
        \pgfmathsetmacro{\cubey}{0.25}
        \pgfmathsetmacro{\cubez}{0.25}
        \draw[kit-orange, fill=kit-orange50] (0,0,0) -- ++(-\cubex,0,0) -- ++(0,-\cubey,0) -- ++(\cubex,0,0) -- cycle;
        \draw[kit-orange, fill=kit-orange50] (0,0,0) -- ++(0,0,-\cubez) -- ++(0,-\cubey,0) -- ++(0,0,\cubez) -- cycle;
        \draw[kit-orange, fill=kit-orange50] (0,0,0) -- ++(-\cubex,0,0) -- ++(0,0,-\cubez) -- ++(\cubex,0,0) -- cycle;
    \end{scope}
    \node at (7.375, -4.3) {\scriptsize linear};

    \begin{scope}[xshift=3.075cm, yshift=1.5cm]
        \pgfmathsetmacro{\cubex}{0.2}
        \pgfmathsetmacro{\cubey}{0.2}
        \pgfmathsetmacro{\cubez}{0.2}
        \draw[kit-blue, fill=kit-blue30] (0,0,0) -- ++(-\cubex,0,0) -- ++(0,-\cubey,0) -- ++(\cubex,0,0) -- cycle;
        \draw[kit-blue, fill=kit-blue30] (0,0,0) -- ++(0,0,-\cubez) -- ++(0,-\cubey,0) -- ++(0,0,\cubez) -- cycle;
        \draw[kit-blue, fill=kit-blue30] (0,0,0) -- ++(-\cubex,0,0) -- ++(0,0,-\cubez) -- ++(\cubex,0,0) -- cycle;
    \end{scope}
    \node[anchor=west] at(3.275, 1.45) {\scriptsize 2D conv, batch norm, ReLU};
    \begin{scope}[xshift=3.075cm, yshift=1.2cm]
        \pgfmathsetmacro{\cubex}{0.2}
        \pgfmathsetmacro{\cubey}{0.2}
        \pgfmathsetmacro{\cubez}{0.2}
        \draw[kit-green, fill=kit-green30] (0,0,0) -- ++(-\cubex,0,0) -- ++(0,-\cubey,0) -- ++(\cubex,0,0) -- cycle;
        \draw[kit-green, fill=kit-green30] (0,0,0) -- ++(0,0,-\cubez) -- ++(0,-\cubey,0) -- ++(0,0,\cubez) -- cycle;
        \draw[kit-green, fill=kit-green30] (0,0,0) -- ++(-\cubex,0,0) -- ++(0,0,-\cubez) -- ++(\cubex,0,0) -- cycle;
    \end{scope}
    \node[anchor=west] at(3.275, 1.15) {\scriptsize max pooling};
    \begin{scope}[xshift=3.075cm, yshift=0.9cm]
        \pgfmathsetmacro{\cubex}{0.2}
        \pgfmathsetmacro{\cubey}{0.2}
        \pgfmathsetmacro{\cubez}{0.2}
        \draw[kit-red, fill=kit-red30] (0,0,0) -- ++(-\cubex,0,0) -- ++(0,-\cubey,0) -- ++(\cubex,0,0) -- cycle;
        \draw[kit-red, fill=kit-red30] (0,0,0) -- ++(0,0,-\cubez) -- ++(0,-\cubey,0) -- ++(0,0,\cubez) -- cycle;
        \draw[kit-red, fill=kit-red30] (0,0,0) -- ++(-\cubex,0,0) -- ++(0,0,-\cubez) -- ++(\cubex,0,0) -- cycle;
    \end{scope}
    \node[anchor=west] at(3.275, 0.85) {\scriptsize average pooling};
    \begin{scope}[xshift=3.075cm, yshift=0.6cm]
        \pgfmathsetmacro{\cubex}{0.2}
        \pgfmathsetmacro{\cubey}{0.2}
        \pgfmathsetmacro{\cubez}{0.2}
        \draw[kit-orange, fill=kit-orange30] (0,0,0) -- ++(-\cubex,0,0) -- ++(0,-\cubey,0) -- ++(\cubex,0,0) -- cycle;
        \draw[kit-orange, fill=kit-orange30] (0,0,0) -- ++(0,0,-\cubez) -- ++(0,-\cubey,0) -- ++(0,0,\cubez) -- cycle;
        \draw[kit-orange, fill=kit-orange30] (0,0,0) -- ++(-\cubex,0,0) -- ++(0,0,-\cubez) -- ++(\cubex,0,0) -- cycle;
    \end{scope}
    \node[anchor=west] at(3.275, 0.55) {\scriptsize fully-connected, softmax};

    \node[anchor=west, align=left] at (8.175, -1.45) {\scriptsize COVID-19\\\scriptsize infection\\\scriptsize binary decision};
\end{tikzpicture}